\documentclass[graybox]{svmult}

\usepackage{mathptmx}       
\usepackage{helvet}         
\usepackage{courier}        
\usepackage{type1cm}        
\usepackage{makeidx}         
\usepackage{graphicx}        
\usepackage{multicol}        
\usepackage[bottom]{footmisc}
\usepackage{adjustbox}
\usepackage{caption}
\usepackage{rotating}

\makeindex             
\font\tabfont=cmr12 at 9pt

\begin{document}

\title*{Fog Computing: A Taxonomy, Survey and Future Directions}
\author{Redowan Mahmud, Ramamohanarao Kotagiri and Rajkumar Buyya}
\institute{Redowan Mahmud \at Cloud Computing and
Distributed Systems (CLOUDS) Laboratory, Department of Computing and
Information System, The University of Melbourne, Parkville VIC 3010, Australia,\\ \email{mahmudm@student.unimelb.edu.au}
\and Ramamohanarao Kotagiri \at Cloud Computing and
Distributed Systems (CLOUDS) Laboratory, Department of Computing and
Information System, The University of Melbourne, Parkville VIC 3010, Australia,\\  
\and Rajkumar Buyya \at Cloud Computing and
Distributed Systems (CLOUDS) Laboratory, Department of Computing and
Information System, The University of Melbourne, Parkville VIC 3010, Australia, \\
}
%
%
\maketitle

\abstract{In recent years, the number of Internet of Things (IoT) devices/sensors has increased to a great extent. To support the computational demand of real-time latency-sensitive applications of largely geo-distributed IoT devices/sensors, a new computing paradigm named "Fog computing" has been introduced. Generally, Fog computing resides closer to the IoT devices/sensors and extends the Cloud-based computing, storage and networking facilities. In this chapter, we comprehensively analyse the challenges in Fogs acting as an intermediate layer between IoT devices/sensors and Cloud datacentres and review the current developments in this field. We present a taxonomy of Fog computing according to the identified challenges and its key features. We also map the existing works to the taxonomy in order to identify current research gaps in the area of Fog computing. Moreover, based on the observations, we propose future directions for research.}

\section{Introduction} \label{Sec.Intro}
Fog computing is a distributed computing paradigm that acts as an intermediate layer in between Cloud datacentres and IoT devices/sensors. It offers compute, networking and storage facilities so that Cloud-based services can be extended closer to the IoT devices/sensors \cite{Intro-Fog_Computing_principles}. The concept of Fog computing was first introduced by Cisco in 2012 to address the challenges of IoT applications in conventional Cloud computing. IoT devices/sensors are highly distributed at the edge of the network along with real-time and latency-sensitive service requirements. Since Cloud datacentres are geographically centralized, they often fail to deal with storage and processing demands of billions of geo-distributed IoT devices/sensors. As a result, congested network, high latency in service delivery, poor Quality of Service (QoS) are experienced \cite{Intro-Theoretical_modelling}.

\begin{figure}[!ht]
\centering 
\captionsetup{justification=centering,margin=2cm}
\includegraphics[width=70mm, height= 80mm]{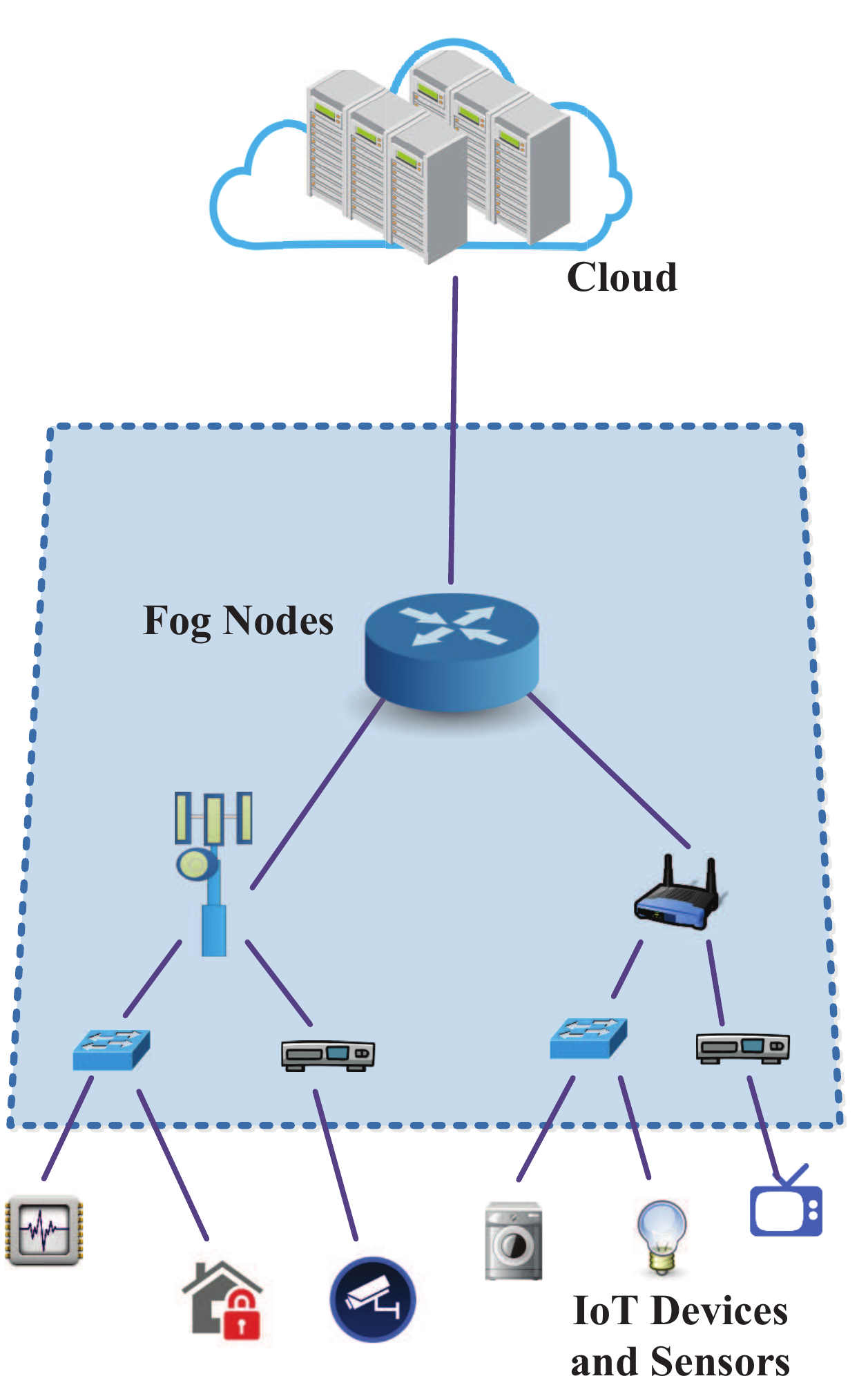}
\caption{Fog Computing Environment}
\label{Fig.System}
\end{figure}
\par Typically, a Fog computing environment is composed of traditional networking components e.g. routers, switches, set top boxes, proxy servers, Base Stations (BS), etc. and can be placed at the closer proximity of IoT devices/sensors as shown in Fig. \ref{Fig.System}. These components are provided with diverse computing, storage, networking, etc. capabilities and can support service-applications execution. Consequently, the networking components enable Fog computing to create large geographical distributions of Cloud-based services. Besides, Fog computing facilitates location awareness, mobility support, real-time interactions, scalability and interoperability \cite{Intro-Fog_computing_role}. Thereby, Fog computing can perform efficiently in terms of service latency, power consumption, network traffic, capital and operational expenses, content distribution, etc. In this sense, Fog computing better meets the requirements with respect to IoT applications compared to a solely use of Cloud computing \cite{Intro-Assessment_Suitability}. 
\par However, the concept of Fog computing is very much similar to the existing computing paradigms. In this chapter, we elaborately discuss the fundamental differences of Fog computing with other computing paradigms. Here, we also analyse different aspects of Fog computing including corresponding resource architecture, service quality, security issues, etc. and review recent research works from the literature. We present a taxonomy based on the key properties and associated challenges in Fog computing. We map the existing works to the taxonomy to identify innovative approaches and limitations in this field. Based on the observations, we also propose potential future directions so that further improvement in Fog computing can be achieved.     
\par The rest of the chapter is organized as follows. In Section 2, we discuss the differences of Fog computing with other related computing approaches. After that, we describe the challenges of Fog computing and propose our taxonomy in Section 3 and  Section 4, respectively. From Section 5 to Section 10, we map the existing research works to the proposed taxonomy. In Section 11, we analyze research gaps and present some promising directions towards future research in Fog computing. Finally, we summarize the findings and conclude the paper.

\section{Related Computing Paradigms} \label{Sec.Similar}

With the origination of Cloud computing, computation technology has entered to a new era. Many computation service providers including Google, Amazon, IBM, Microsoft, etc. are currently nurturing this popular computing paradigm as a utility. They have enabled cloud based services such as Infrastructure as a Service (IaaS), Platform as a Service (PaaS), Software as a Service (SaaS), etc. to handle numerous enterprise and educational related issues simultaneously. However, most of the Cloud datacentres are geographically centralized and situated far from the proximity of the end devices/users. As a consequence, real-time and latency-sensitive computation service requests to be responded by the distant Cloud datacentres often endure large round-trip delay, network congestion, service quality degradation, etc. To resolve these issues besides centralized Cloud computing, a new concept named "Edge computing" has recently been proposed \cite{EdgeComputing}.
             
\par The fundamental idea of Edge computing is to bring the computation facilities closer to the source of the data. More precisely, Edge computing enables data processing at the edge network \cite{newEdge}. Edge network basically consists of end devices (e.g. mobile phone, smart objects, etc.), edge devices (e.g. border routers, set-top boxes, bridges, base stations, wireless access points etc.), edge servers, etc. and these components can be equipped with necessary capabilities for supporting edge computation. As a localized computing paradigm, Edge computing provides faster responses to the computational service requests and most often resists bulk raw data to be sent towards core network. However, in general Edge computing does not associate IaaS, PaaS, SaaS and other cloud based services spontaneously and concentrate more towards the end devices side \cite{EdgeComputing-1}. 

\par Taking the notion of Edge and Cloud computing into account, several computing paradigms have already been introduced in computation technology. Among them Mobile Edge Computing (MEC), Mobile Cloud Computing (MCC) are considered as the potential extensions of Cloud and Edge computing.  

\par As an edge-centric computing paradigm, MEC has already created significant buzz in the research domain. MEC has been regarded as one of the key enablers of modern evolution of cellular base stations. It offers edge servers and cellular base staions to be operated combinedly \cite{EdgeComputing1}. MEC can be either connected or not connected with distant Cloud datacentres. Thus along with end mobile devices, MEC supports 2 or 3 tire hierarchical application deployment in the network \cite{Differences}. Besides, MEC targets adaptive and faster initiation of cellular services for the customers and enhances network efficiency. In recent times, significant improvement in MEC has been made so that it can support 5G communications. Moreover, it aims at flexible access to the radio network information for content distribution and application development \cite{EdgeComputing2} \cite{EdgeComputing3}.

\par MCC is another recent trend in computation. Due to the proliferation of smart mobile devices, nowadays end users prefer to run necessary applications in their handheld devices rather than traditional computers. However, most of the smart mobile devices are subjected to energy, storage and computational resource constraints \cite{MCC1}. Therefore, in critical scenarios, it is more feasible to execute compute intensive applications outside the mobile devices compared to execute those applications locally. In this case, MCC provides necessary computational resources to support such remote execution of offloaded mobile applications in closer proximity of end users \cite{MCC2} \cite{MCC3}. In MCC, most often light-weight cloud servers named cloudlet \cite{cloudlets1} are placed at the edge network. Along with end mobile devices and Cloud datacentres, cloudlets develop a 3 tire hierarchical application deployment platform for rich mobile applications \cite{Differences}. In brief, MCC combines cloud computing, mobile computing and wireless communication to enhance Quality of Experience (QoE) of mobile users and creates new business opportunities for both network operators and cloud service providers.

\par
\begin{figure}[h]
\centering 
\captionsetup{justification=centering,margin=2cm}
\includegraphics[width=130mm, height= 120mm]{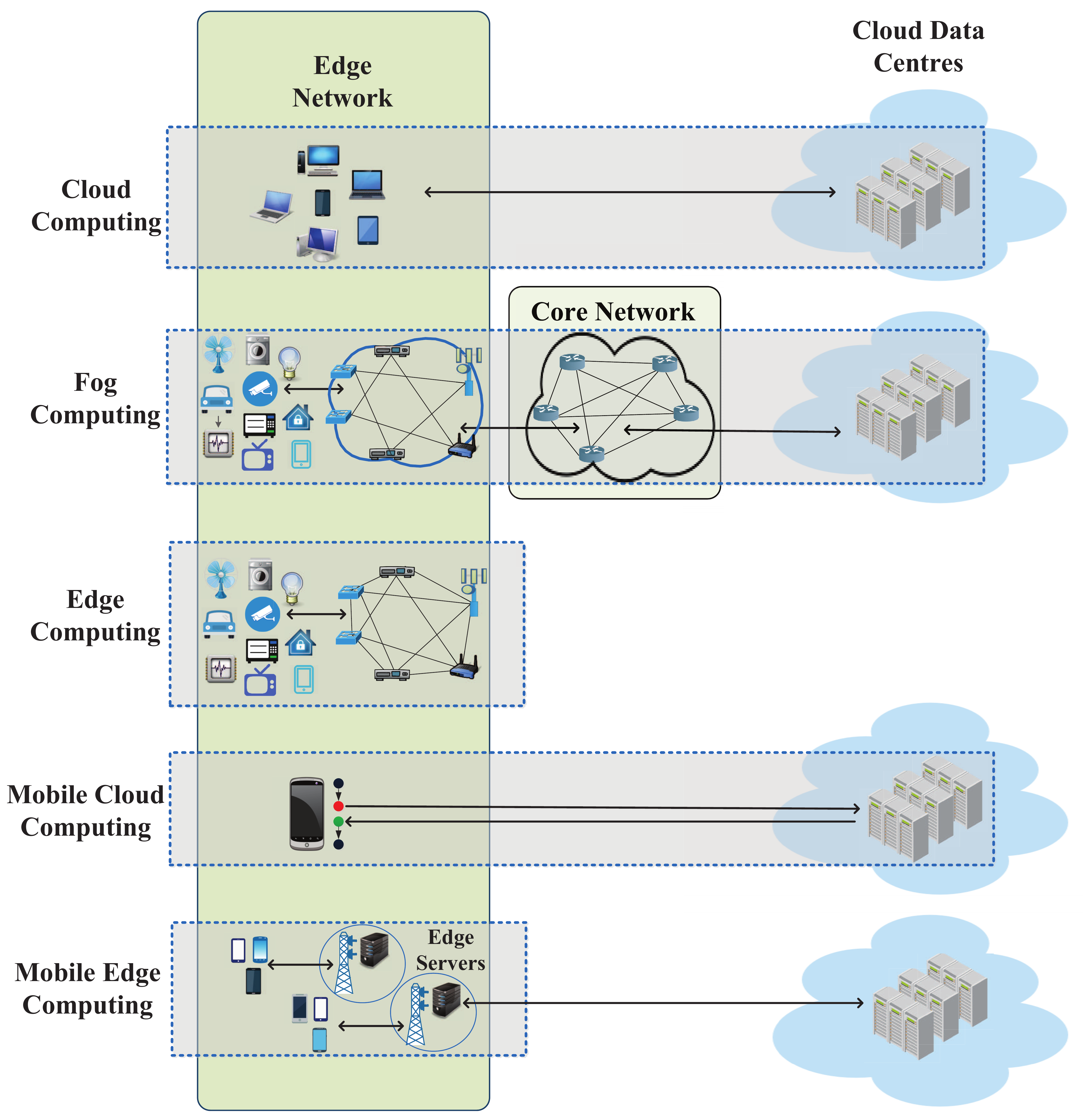}
\caption{Computation domain of Cloud, Fog, Edge, Mobile Cloud and Mobile Edge computing}
\label{Fig.Differences}
\end{figure}

 Like MEC and MCC, Fog computing can also enable edge computation. However, besides edge network, Fog computing can be expanded to the core network as well \cite{Intro-Fog_computing_role}. More precisely, both edge and core networking (e.g.  core routers, regional servers, wan switches, etc.) components can be used as computational infrastructure in Fog computing (Fig. \ref{Fig.Differences}). As a consequence, multi-tire application deployment and service demand mitigation of huge number of IoT devices/sensors can easily be observed through Fog computing. In addition, Fog computing components at the edge network can be placed closer to the IoT devices/sensors compared to cloudlets and cellular edge servers. As IoT devices/sensors are densely distributed and require real-time responses to the service requests, this approach enables IoT data to be stored and processed within the vicinity of IoT device/sensors. As a result, service delivery latency for real-time IoT applications will be minimized to a great extent. Unlike Edge computing, Fog computing can extend cloud based services like IaaS, PaaS, SaaS, etc. to the edge of the network as well. Due to the aforementioned features, Fog computing is considered as more potential and well structured for IoT compared to other related computing paradigms.

\section{Challenges in Fog computing} \label{Sec.Challenges}

Fog computing is considered as the promising extension of Cloud computing paradigm to handle IoT related issues at the edge of network. However, in Fog computing, computational nodes are heterogeneous and distributed. Besides, Fog based services have to deal with different aspects of constrained environment. Assurance of security is also predominant in Fog computing. 
\par Analysing the features of Fog computing from structural, service oriented and security perspectives, the challenges in this field can be listed as follows.

\begin{itemize}
\item \textbf{Structural issues.}
\begin{itemize}

\item Different components from both edge and core network can be used as potential Fog computing infrastructure. Typically these components are equipped with various kinds of processors but are not employed for general purpose computing. Provisioning the components with general purpose computation besides their traditional activities will be very challenging. 

\item Based on operational requirements and execution environment, the selection of suitable nodes, corresponding resource configuration and places of deployment are vital in Fog as well.

\item In Fog computing, computational nodes are distributed across the edge network and can be virtualized or shared. In this case, identification of appropriate techniques, metrics, etc. for inter-nodal collaboration and efficient resource provisioning are important. 
\item The structural orientation of Fog computing is compatible for IoT. However, competency assurance of Fog computing in other networking systems such as Content Distribution Network (CDN), vehicular network, etc. will be very challenging.                 
\end{itemize} 
\end{itemize} 

\begin{itemize}
\item \textbf{Service oriented.}
\begin{itemize}
\item Not all Fog nodes are resource enriched. Therefore, large scale applications development in resource constrained nodes are not quite easy compared to conventional datacentres. In this case, potential programming platform for distributed applications development in Fog are required to be introduced. 
\item Policies to distribute computational tasks and services among IoT devices/ sensors, Fog and Cloud infrastructures are required to be specified. Data visualization through web-interfaces are also difficult to design in Fog computing.
\item In Fog computing, the Service Level Agreement (SLA) is often affected by many factors such as, service cost, energy usage, application characteristics, data flow, network status etc. Therefore, on a particular scenario, it is quite difficult to specify the service provisioning metrics and corresponding Service Level Objectives (SLOs). Besides, it is highly required to retain the fundamental QoS of the Fog nodes for which they are actually designed. 

\end{itemize} 
\end{itemize}   

\begin{itemize}
\item \textbf{Security aspects.}
\begin{itemize}
\item Since Fog computing is designed upon traditional networking components, it is highly vulnerable to security attacks. 
\item Authenticated access to services and maintenance of privacy in a largely distributed paradigm like Fog computing are hard to ensure. 
\item Implementation of security mechanisms for data-centric integrity can affect the QoS of Fog computing to a great extent.                 
\end{itemize} 
\end{itemize} 

In addition to aforementioned challenges service scalability, end users QoE, context-awareness, mobility support are very crucial performance indicator for Fog computing and very difficult to deal with in real-time interactions.

\section{Taxonomy} \label{Sec.Taxonomy}

Fig. 3 represents our proposed taxonomy for Fog computing. Based on the identified challenges from Section \ref{Sec.Challenges}, the taxonomy provides a classification of the existing works in Fog computing. More precisely, the taxonomy highlights the following aspects in Fog computing.

\begin{itemize}
\item[--] Fog Nodes Configuration.
\par The computational nodes with heterogeneous architecture and configurations that are capable to provide infrastructure for Fog computing at the edge of the network.  
\end{itemize}

\begin{itemize}
\item[--] Nodal Collaboration.
\par The techniques for managing nodal collaboration among different Fog nodes within the edge network. 
\end{itemize}
 
\begin{itemize}
\item[--] Resource/Service Provisioning Metric.
\par The factors that contribute to provision resources and services efficiently under different constraints. 
\end{itemize}

\begin{itemize}
\item[--] Service Level Objectives.
\par The SLOs that have been attained by deploying Fog computing as an intermediate layer between Cloud datacentres and end devices/sensors.     
\end{itemize}

\begin{itemize}
\item[--] Applicable Network System.
\par The different networking systems where Fog computing has been introduced as extension of other computing paradigms.   
\end{itemize}
 
\begin{itemize}
\item[--] Security Concern.
\par The security issues that have been considered in Fog computing on different circumstances.    
\end{itemize}

Proposed system and corresponding solutions in the existing works generally covers different categories of the taxonomy. However, as this taxonomy is designed based on the associated features of Fog computing, it does not reflect the relative performance of the proposed solutions. Actually, the reviewed existing works considers diverse execution environment, networking topology, application characteristics, resource architecture, etc. and targets different challenges. Therefore, identification of the best approach for Fog computing in terms of structural, service and security aspects is very difficult.

\par In the following sections (from Section 5 to Section 10), we map the existing works in Fog computing to our proposed taxonomy and discuss different facts in detail.

\begin{figure}[!ht]
\vspace{5mm}
\captionsetup{justification=centering,margin=2cm}
  \begin{adjustbox} {addcode={\begin{minipage}{\width}}{\caption{%
      Taxonomy of Fog Computing.
      }\end{minipage}},rotate=270,center}
      \includegraphics[width=177mm, height= 150mm]{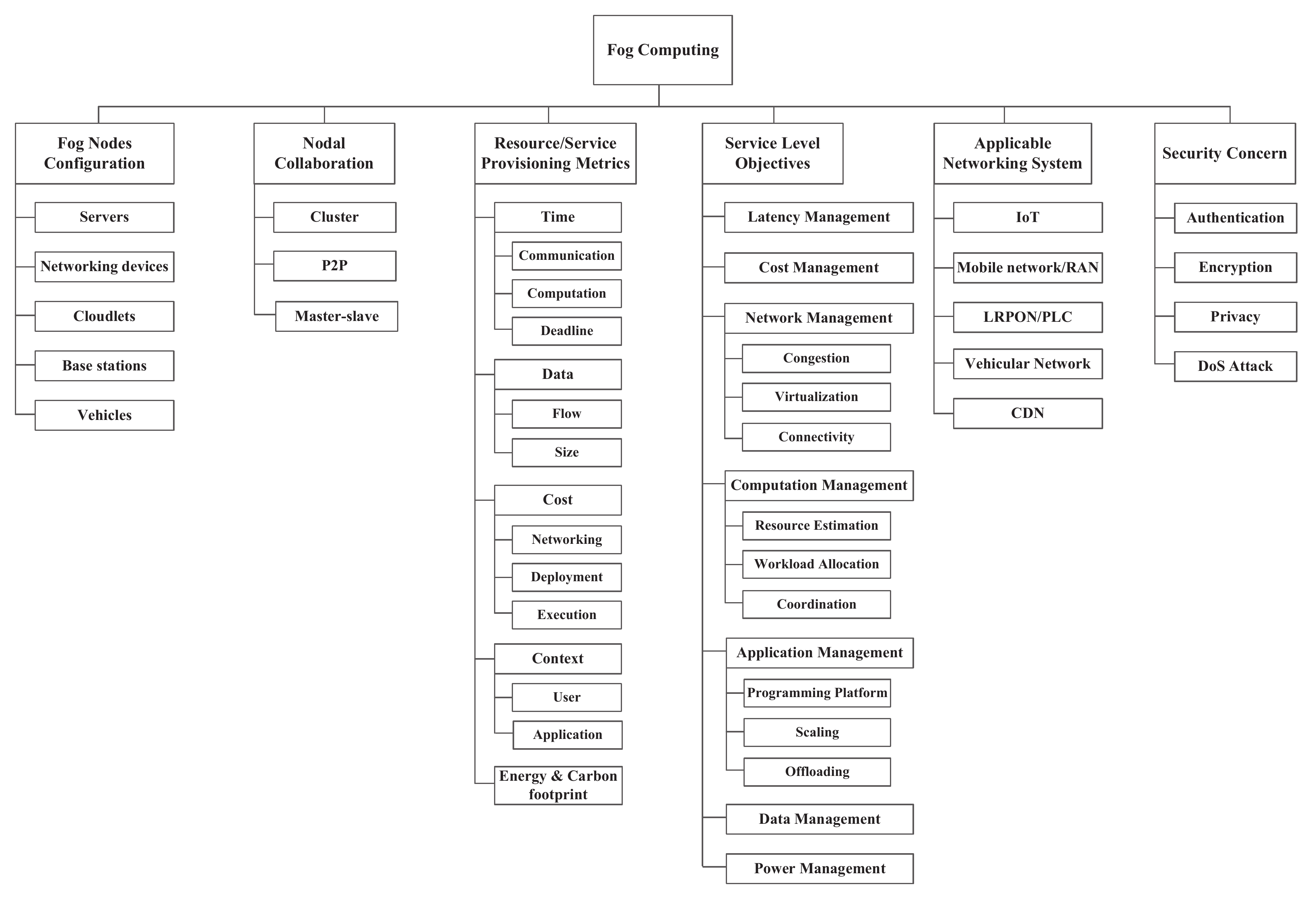} 
  \end{adjustbox}
\end{figure}

\section{Fog Nodes Configuration} \label{Sec.FogNodes}

Five types of Fog nodes and their configurations have been mentioned in the literature: namely servers, networking devices, cloudlets, base stations, vehicles.  

\subsection{Servers}

The Fog servers are geo-distributed and are deployed at very common places for example; bus terminals, shopping centres, roads, parks, etc. Like light-weight Cloud servers, these Fog servers are virtualized and equipped with storage, compute and networking facilities. There are many works that have considered Fog servers as main functional component of Fog computing.
\par In some papers based on the physical size, Fog servers are termed as micro servers, micro datacentres (Lee et al. \cite{Archi-WSN}, Aazam et al. \cite{Model-Micro_Datacenter_Based}), nano servers (Jalali et al. \cite{fog_computing_may_help}), etc. whereas other works categorize Fog servers based on their functionalities like cache servers (Zhu et al. \cite{improving_web_sites_performance}), computation servers, storage servers (Zeng et al. \cite{joint_optimization_of_task}), etc. Server based Fog node architecture enhances the computation and storage capacity in Fog computing. However, it limits the pervasiveness of the execution environment.      

\subsection{Networking Devices}

Devices like gateway routers, switches, set-top boxes, etc. besides their traditional networking activities (routing, packet forwarding, analog to digital signal conversions, etc.) can act as potential infrastructure for Fog computing. In some existing works, the networking devices are designed with certain system resources including data processors, extensible primary and secondary memory, programming platforms, etc. (Hong et al. \cite{mobile_fog_a_programming}, Nazmudeen et al. \cite{improved_throughput_for_power}).
\par In other works, apart from conventional networking devices, several dedicated networking devices like \textit{Smart gateways} (Aazam et al. \cite{Archi-Smart_Gateway_Based}), \textit{IoT Hub} (Cirani et al. \cite{Archi-IoT_hub}) have been introduced as Fog nodes. Distributed deployment of networking devices helps Fog computing to be ubiquitous although physical diversity of the devices significantly affects service and resource provisioning.            

\subsection{Cloudlets}

Cloudlets are considered as micro-cloud and loacted at the middle layer of end device, cloudlet, and Cloud hierarchy. Basically cloudlets have been designed for extending Cloud based services towards mobile device users and can complement MCC \cite{MCC1}. In several works (Dsouza et al. \cite{policy_driven_security_management}, Cardellini et al. \cite{on_qos_aware_scheduling}), cloudlets are mentioned as Fog nodes. Cloudlet based Fog computing are highly virtualized and can deal with large number of end devices simultaneously.  
\par In some cases, due to structural constraints, cloudlets even after deploying at the edge act as centralized components. In this sense, the limitations of centralized computation still remain significant in Fog computing which resist to support IoT.

\subsection{Base Stations}  

Base stations are very important components in mobile and wireless networks for seamless communication and data signal processing. In recent works, traditional base stations equipped with certain storing and computing capabilities are considered suitable for Fog computing (Yan et al. \cite{user_access_mode_selection}, Gu et al. \cite{Obj-Cost_Efficient}). Like traditional base stations, Road Side Unit (RSU) (Truong et al. \cite{software_defined_networking_based}) and small cell access points (Oueis et al. \cite{small_cell_clustering_for}), etc. can also be used as potential Fog nodes. 

\par Base stations are preferable for Fog based extension of Cloud Radio Access Network (CRAN), Vehicular Adhoc Network (VANET), etc. However, formation of a dense Fog environment with base stations is subjected to networking interference and high deployment cost.  

\subsection{Vehicles} 

Moving or parked vehicles at the edge of network with computation facilities can serve as Fog nodes (Hou et al. \cite{vehicular_fog_computing_a}, Ye et al. \cite{scalable_fog_computing_with}). Vehicles as Fog nodes can form a distributed and highly scalable Fog environment. However, the assurance of privacy and fault tolerance along with desired QoS maintenance will be very challenging in such environment.      

\section{Nodal Collaboration} \label{Sec.Collaboration}

Three techniques (cluster, peer to peer and master-slave) for nodal collaboration in Fog computing have been specified in the literature.

\subsection{Cluster}

Fog nodes residing at the edge can maintain a collaborative execution environment by forming cluster among themselves. Clusters can be formed either based on the homogeneity of the Fog nodes (Cardellini et al. \cite{on_qos_aware_scheduling}) or their location (Hou et al. \cite{vehicular_fog_computing_a}). Computational load balancing (Oueis et al. \cite{the_fog_balancing_load}) and functional sub-system development (Faruque et al. \cite{energy_management_as_a}) can also be given higher priority while forming cluster among the nodes. 

\par Cluster based collaboration is effective in exploiting capabilities of several Fog nodes simultaneously. However, static clusters are difficult to make scalable in runtime and dynamic formation of clusters largely depends on the existing load and the availability of Fog nodes. In both cases networking overhead plays a vital role.

\subsection{Peer to Peer}

In Fog computing Peer to Peer (P2P) collaboration among the nodes is very common. P2P collaboration can be conducted in both hierarchical (Hong et al. \cite{mobile_fog_a_programming}) and flat order (Shi et al. \cite{combining_mobile_and_fog}). Besides based on proximity, P2P collaboration between Fog nodes can be classified as home, local, non-local, etc. (Jalali et al. \cite{fog_computing_may_help}). Through P2P collaboration not only processed output of one node appears as input to another node (Giang et al. \cite{developing_iot_applications_in}) but also virtual computing instances are shared between the nodes (Dsouza et al. \cite{policy_driven_security_management}). 
\par Augmentation of Fog nodes in P2P collaboration is quite simple and nodes can be made reusable. However, reliability and access control related issues are predominant in P2P nodal collaboration.         

\subsection{Master-Slave}

In several works, master-slave based nodal collaboration has been mentioned elaborately. Through master-slave based collaboration generally a master Fog node controls functionalities, processing load, resource management, data flow, etc. of underlying slave nodes (Lee et al. \cite{Archi-WSN}).
\par Besides, master-slave based approach along with cluster and P2P based nodal interactions can form a hybrid collaborative network within the Fog computing environment (Nazmudeen et al. \cite{improved_throughput_for_power}, Truong et al. \cite{software_defined_networking_based}). However, in real-time data processing due to this kind of functional decomposition, the master and the slave Fog nodes require high bandwidth to communicate with each other.             

\section{Resource/Service Provisioning Metrics} \label{Sec.Metrics}

In existing literature of Fog computing many factors including time, energy, user-application context, etc. have been found playing important roles in resource and service provisioning. 

\subsection{Time} 

In Fog computing paradigm, time is considered as one of the important factors for efficient resource and service provisioning.

\par \textbf{Computation time} refers to the required time for task execution. Computation time of an application largely depends on resource configuration where the application is running or the task has been scheduled (Zeng et al. \cite{joint_optimization_of_task}) and can be changed according to the existing load. Besides, task computation time helps to identify the active and idle periods of different applications which significantly influences resource and power management in Fog (Jalali et al. \cite{fog_computing_may_help}).
   
\par \textbf{Communication time} basically defines the networking delay to exchange data elements in Fog computing environment. In the literature it has been discussed in 2 folds: End device/sensors to Fog nodes (Intharawijitr et al. \cite{analysis_of_fog_model}), Fog nodes to Fog nodes (Zeng et al. \cite{joint_optimization_of_task}). Required communication time reflects the network context which assists selection of suitable Fog nodes for executing tasks.    

\par \textbf{Deadline} specifies the maximum delay in service delivery that a system can tolerate. In some papers deadline satisfied task completion has been considered as important parameter for measuring QoS of the system (Oueis et al. \cite{small_cell_clustering_for}, Ye et al. \cite{scalable_fog_computing_with}). Basically service delivery deadline plays a significant role in characterizing latency sensitive and latency tolerant applications. 

\par In addition, the impact of other time based metrics like data sensing frequency of end device/sensors, service access time in multi-tenant architecture, expected service response time, etc. can be investigated for efficient service and resource provisioning in Fog computing.

\subsection{Data}

Among data-centric metrics, input data size and data flow characteristics are found very common in Fog computing literature. 

\par \textbf{Data size} points to the amount of data that has to be processed through Fog computing. In several works, data size has been discussed in respect of computational space requirements of the requests (Shi et al. \cite{combining_mobile_and_fog}). Besides, bulk data collected from distributed devices/sensors can contain the features of Big Data (Nazmudeen et al. \cite{improved_throughput_for_power}) as well. In this case, provisioning resource and service according to the data load can be an effective approach. Moreover, data size plays an important role in making decision about either local or remote processing of the corresponding computaional tasks (Peng et al. \cite{fog_computing_based_radio}).

\par \textbf{Data flow} defines the characteristics of data transmission. Data flow through out the Fog computing environment can be event driven (Lee et al. \cite{Archi-WSN}) or real time (Giang et al. \cite{developing_iot_applications_in}) and can influence resource and associated service provisioning to a great extent. Besides, sudden change in data flow sometimes promotes dynamic load balancing among the nodes (Oueis et al. \cite{the_fog_balancing_load}). 

\par Moreover, the effectiveness of heterogeneous data architecture, data semantic rules, data integrity requirements can also be studied for resource and service provisioning in Fog computing. 

\subsection{Cost}
In certain cases, cost related factors form both service providers and users perspective become very influential in Fog resource and service provisioning. 

\par \textbf{Networking cost} in Fog computing environment is directly related to the bandwidth requirements and associated expenses. In several works data uploading cost from end devices/sensors and inter-nodal data sharing cost have been considered as the elements of networking cost (Gu et al. \cite{Obj-Cost_Efficient}) whereas in other works, experienced network latency due to bandwidth issues has been termed as networking cost (Hassan et al. \cite{help_your_mobile_applications}).

\par \textbf{Deployment cost} is basically associated with the infrastructure placement related expenses in Fog computing environment. In some papers cost-effective infrastructure deployment has been considered supportive for efficient resource and service provisioning. Infrastructure deployment cost can be discussed in terms of both placing Fog nodes in the network (Zhang et al. \cite{infrastructure_deployment_and_optimization}) and creating virtual computing instances in Fog nodes (Gu et al. \cite{Obj-Cost_Efficient}). 

\par \textbf{Execution cost} refers to the computational expenses of Fog nodes while running applications or processing tasks. Although in other computing paradigms execution cost is widely used in resource provisioning and billing, in Fog computing this metric has been used in very few works. In these works total execution cost has been calculated in terms of task completion time and per unit time resource usage cost (Hassan et al. \cite{help_your_mobile_applications}). 

\par In addition to aforementioned costs, expenses related to security safeguards, the maximum price that an user is willing to pay for a service, migration costs can also be considered for resource and service provisioning in Fog computing.       

\subsection{Energy Consumption and Carbon Footprint}

In several works, energy related issues have been given higher priority in provisioning Fog resources and services. The energy consumption of all devices in home-based Fog computing environment (Faruque et al. \cite{energy_management_as_a}) and the energy-latency tradeoff in different phases of Fog-Cloud interaction (Deng et al. \cite{Obj-Delay_Power}) have been highlighted widely in these works. In another work, carbon emission rate of different nodes in respect of per unit energy usage have been considered for resource provisioning purposes (Do et al. \cite{Obj-carbon_footprint}).
\par As end devices/sensors are energy-constrained, energy aspects of end components for example residual battery lifetime, energy-characteristics of communication medium can also be investigated in provisioning Fog resources.  

\subsection{Context}

Context refers to situation or condition of a certain entity in different circumstances. In Fog based research works user and application level context have been discussed for resource and service provisioning. 

\par \textbf{User} context such as user characteristics, service usage history, service relinquish probability, etc. can be used for allocating resources for that user in future (Aazam et al. \cite{Model-IoT_trace_base}). Users service feedback for example Net Promoter Score (NPS) and user requirements (Datta et al. \cite{fog_computing_architecture_to}) can also be used for service and resource provisioning purposes (Aazam et al. \cite{mefore_qoe_based_resource}). In other works users density (Yan et al. \cite{user_access_mode_selection}), mobility (Hou et al. \cite{vehicular_fog_computing_a}) and network status (Zhu et al. \cite{improving_web_sites_performance}) have also been considered for service provisioning.                                  

\par \textbf{Application} context can be considered as operational requirements of different applications. Operational requirements includes task processing requirements ( CPU speed, storage, memory) (Aazam et al. \cite{Archi-Smart_Gateway_Based}, Truong et al. \cite{software_defined_networking_based}, Gazis et al. \cite{components_of_fog_computing}), networking  requirements (Cirani et al. \cite{Archi-IoT_hub}, Dsouza et al. \cite{policy_driven_security_management}), etc. and can affect resource and service provisioning. In other works current task load of different applications  (Cardellini et al. \cite{on_qos_aware_scheduling}, Shi et al. \cite{combining_mobile_and_fog}) have also been considered as application context.

\par Moreover, contextual information in Fog computing can be discussed in terms of execution environment, nodal characteristics, application architecture, etc. and along with the other contexts they can play vital roles in provisioning resource and services. Therefore, it is essential to investigate the impact of every contextual information in-detail.       

\section{Service Level Objectives}
In existing literature, several unique Fog node architecture, application programming platform, mathematical model and optimization technique have been proposed to attain certain SLOs. Most of the attained SLOs are management oriented and cover latency, power, cost, resource, data, application, etc. related issues. 

\subsection{Latency Management}

Latency management in Fog computing basically resists the ultimate service delivery time from surpassing an accepted threshold. This threshold can be the maximum tolerable latency of a service request or applications QoS requirement.

\par To ensure proper latency management, in some works efficient initiation of nodal collaboration has been emphasized so that the computation tasks through the collaborated nodes can be executed within the imposed latency constraints (Oueis et al. \cite{small_cell_clustering_for}). In another work, computation task distribution between the client and Fog nodes have been conducted with a view to minimizing the total computation and communication latency of service requests (Zeng et al. \cite{joint_optimization_of_task}).     

\par Besides, in another work architecture of low-latency Fog network has been proposed for latency management (Intharawijitr et al. \cite{analysis_of_fog_model}). The basic intention of this work is to select that node from the Fog network which provides lowest latency in service delivery.

\subsection{Cost Management} 

Cost management in Fog computing can be discussed in terms of Capital Expenses (CAPEX) and Operating Expenses (OPEX).

\par The main contributor of CAPEX in Fog computing is the deployment of cost of distributed Fog nodes and their associated networks. In this case, suitable placement and optimized number of Fog nodes play a significant role in minimizing the CAPEX in Fog computing. Investigating this issue, a Fog computing network architecture has been proposed in (Zhang et al. \cite{infrastructure_deployment_and_optimization}) that minimizes the total CAPEX in fog computing by optimizing the places and numbers of Fog node deployment.      

\par In another work (Gu et al. \cite{Obj-Cost_Efficient}), Fog nodes have been considered as virtualized platforms for launching VMs. Execution of data processing operations in these VMs are not free of cost and the cost can be varied from provider to provider. Therefore, it is feasible to exploit cost-diversity of different Fog nodes/ providers for minimizing OPEX in Fog computing. In respect to this fact, a solution to find suitable set of Fog nodes for hosting VMs has been proposed in that paper which aims to minimize the OPEX in Fog computing.

\subsection{Network Management} 

Network management in Fog computing includes core-network congestion control, support for Software Define Network (SDN)/ Network Function Virtualization (NFV), assurance of seamless connectivity, etc.

\par \textbf{Network congestion} mainly occurs due to increasing overhead on the network. As in IoT, end devices/sensors are highly distributed across the edge, simultaneous interactions of end components with Cloud datacentres can increase the overhead on the core network to a great extent. In such case network congestion will occur and degrade the performance of the system. Taking cognizance of this fact, in (Aazam et al. \cite{Archi-Smart_Gateway_Based}) a layered architecture of Fog node has been proposed that provides local processing of the service requests. As a consequence, despite of receiving bulk service requests, Clouds get consized version of the requests which contribute less to the network congestion. 

\par \textbf{Virtualization} of conventional networking system has already drawn significant research attention. SDN is considered as one of the key enablers of virtualized network. SDN is a networking technique that decouples the control plane from networking equipment and implements in software on separate servers. One of the important aspects of SDN is to provide support for NFV. Basically, NFV is an architectural concept that virtualizes traditional networking functions (network address translation (NAT), firewalling, intrusion detection, domain name service (DNS), caching, etc.) so that they can be executed through software. In Cloud based environment SDN and NFV is quite influencing due to their wide range of services. Being motivated by this, in several research works (Lee et al. \cite{Archi-WSN}, Truong et al. \cite{software_defined_networking_based}, Peng et al. \cite{fog_computing_based_radio}) new network structures of Fog computing have been proposed to enable SDN and NFV. 

\par \textbf{Connectivity} ensures seamless communication of end devices with other entities like Cloud, Fog, Desktop computers, Mobile devices, end devices, etc. despite of their physical diversity. As a consequence, resource discovery, maintenance of communication and computation capacity become easier within the network. Several works in Fog computing have already targeted this issue and proposed new architecture of Fog nodes e.g \textit{IoT Hub} (Cirani et al. \cite{Archi-IoT_hub}) and Fog networking e.g \textit{Vehicular Fog Computing} (Hou et al. \cite{vehicular_fog_computing_a}) for connectivity management and resource discovery. Besides, for secured connectivity among the devices a policy driven framework has also been developed for Fog computing (Dsouza et al. \cite{policy_driven_security_management}).               
  
\subsection{Computation Management} 

Among the attained SLOs, assurance of proper computational resource management in Fog computing is very influential. Fog computing resource management includes resource estimation, workload allocation, resource coordination, etc.

\par \textbf{Resource estimation} in Fog computing helps to allocate computational resources according to some policies so that appropriate resources for further computaton can be allocated, desired QoS can be achieved and accutare service price can be imposed. In existing literature, resource estimation policies are developed in terms of user characteristics , experienced QoE, features of service accessing devices, etc. (Aazam et al. \cite{Model-Micro_Datacenter_Based}\cite{Model-IoT_trace_base}\cite{mefore_qoe_based_resource}).

\par \textbf{Workload allocation} in Fog computing should be done in such a way so that utilization rate of resources become maximized and longer computational idle period get minimized. More precisely, balanced load on different components is ensured. In a Fog based research work (Zeng et al. \cite{joint_optimization_of_task}), scheduling based workload allocation policy has been introduced to balance computation load on Fog nodes and client devices. As a consequence overhead on both parts become affordable and enhance QoE. In another work (Deng et al. \cite{Obj-Delay_Power}) a workload allocation framework has been proposed that balances delay and power consumption in Fog-Cloud interaction.   

\par \textbf{Coordination} among different Fog resources is very essential as they are heterogeneous and resource constrained. Due to decentralized nature of Fog computing, in most cases large scale applications are distributively deployed in different Fog nodes. In such scenarios without proper co-ordination of Fog resources, attainment of desired performance will not be very easy. Considering this fact, in (Giang et al. \cite{developing_iot_applications_in}) a directed graph based resource co-ordination model has been proposed for Fog resource management.

\subsection{Application Management} 

In order to ensure proper application management in Fog computing, efficient programming platforms are very essential. Besides the scalability and computation offloading facilities also contribute significantly in application management. 

\par \textbf{Programming platform} provides necessary components such as interfaces, libraries, run-time environment, etc. to develop, compile and execute applications. Due to dynamic nature of Fog computing, assurance of proper programming support for large-scale applications is very challenging.   In order to overcome this issue, a new programming platform named \textit{Mobile Fog} (Hong et al. \cite{mobile_fog_a_programming}) has been introduced. \textit{Mobile Fog} offers simplified abstraction of programming models for developing large-scale application over heterogeneous-distributed devices. In another paper (Giang et al. \cite{developing_iot_applications_in}), besides coordinating resources during applications execution, a programming platform based on distributed data flow approach has also been designed for application development in Fog computing.        

\par \textbf{Scaling} points to the adaptation capability of applications in retaining their service quality even after proliferation of application users and unpredictable events. Sacling techniques can also be applied in application scheduling and users service access. To support scalable scheduling of data stream applications, architecture of a QoS-aware self adaptive scheduler (Cardellini et al. \cite{on_qos_aware_scheduling}) has been recently proposed in Fog computing. This scheduler can scale applications with the increasing of both users and resources and does not ask for global information about the environment. Moreover, due to self-adaptive capability of the scheduler, automatic reconfiguration of the resources and placement of applications in a distributed fashion become easier. Besides, based on distance, location and QoS requirements of the service accessing entities, an adaptive technique for users service access mode selection has also been proposed in Fog computing (Yan et al. \cite{user_access_mode_selection}).   

\par \textbf{Offloading} techniques facilitate resource constrained end devices in sending their computational tasks to some resource-enriched devices for execution. Computational offloading is very common in mobile cloud environment. However, recently, as a part of compatibility ehnacement of Fog computing for other networking systems, computation offloading support for mobile applications in Fog computing have been emphasized in several papers. In these papers offloading techniques have been discussed in terms of both distributed computation of mobile applications (Shi et al. \cite{combining_mobile_and_fog}) and resources availability (Hassan et al. \cite{help_your_mobile_applications}).        

\subsection{Data Management}

Data management is another important SLO that is highly required to be achieved for efficient performance of Fog computing. In different research works data management in Fog computing has been discussed from different perspectives. In (Aazam et al. \cite{Archi-Smart_Gateway_Based}, Datta et al. \cite{fog_computing_architecture_to}) initiation of proper data analytic services and resource allocation for data pre-processing have been focused for data management policy in Fog computing. Besides, low-latency aggregation of data coming from distributed end devices/sensors can also be considered for efficient data management (Nazmudeen et al. \cite{improved_throughput_for_power}). Moreover, the storage capability of end devices/sensors are not so reach. In this case, storage augmentation in Fog computing for preserving data of end entities can be very influential. Therefore, in (Hassan et al. \cite{help_your_mobile_applications}) besides application management, storage expansion in Fog computing for mobile devices have also been discussed as integral part of data management.               

\subsection{Power Management}

Fog computing can be used as an effective platform for providing power management as a service for different networking systems. In (Faruque et al. \cite{energy_management_as_a}), a service platform for Fog computing has been proposed that can enable power management in home based IoT network with customized consumer control. Additionally, Fog computing can manage power usage of centralized Cloud data centres in certain scenarios. Power consumed by Cloud datacentres largely depends on type of running applications. In this case, Fog computing can complement Cloud datacentres by providing infrastructure for hosting several energy-hungry applications. As a consequence energy consumption in Cloud datacentres will be miniized that eventually ensures proper power management for Cloud datacentres (Jalali et al. \cite{fog_computing_may_help}). Moreover, by managing power in Fog computing emission of carbon footprint can also be controlled (Do et al. \cite{Obj-carbon_footprint}). 

\section{Applicable Network System}

Fog computing plays a significant role in IoT. However, in recent research works the applicability of Fog computing in other networking systems (mobile network, content distribution network , radio access network, vehicular network, etc.) have also been highlighted.  

\subsection{Internet of Things}

In IoT, every devices are interconnected and able to exchange data among themselves. IoT environment can be described from different perspectives. Besides specifying IoT as a network for device to device interaction (Hong et al. \cite{mobile_fog_a_programming}, Cirani et al. \cite{Archi-IoT_hub}, Giang et al. \cite{developing_iot_applications_in}), in several Fog based research works this interaction have been classified under industry (Gazis et al. \cite{components_of_fog_computing}) and home (Faruque et al. \cite{energy_management_as_a}) based execution environment. Moreover, Wireless Sensors and Actuators Network (Lee et al. \cite{Archi-WSN}), Cyber-Physical Systems (Gu et al. \cite{Obj-Cost_Efficient}), Embedded system network (Zeng et al. \cite{joint_optimization_of_task}), etc. have also been considered as different forms of IoT while designing system and service models for Fog computing.         

\subsection{Mobile Network / Radio Access Network}

Mobile network is another networking system where applicability of Fog computing has been explored through several research works. Basically, in these works much emphasize has been given on investigating the compatibility of Fog computing in 5G mobile networking (Oueis et al. \cite{small_cell_clustering_for} \cite{the_fog_balancing_load}, Intharawijitr et al. \cite{analysis_of_fog_model}). 5G enables higher speed communication, signal capacity and much lower latency in service delivery compared to existing cellular systems. Besides 5G, Fog computing can also be applied in other mobile networks like 3G, 4G, etc. Moreover, in another work (Deng et al. \cite{Obj-Delay_Power}), power-delay tradeoff driven workload allocation in Fog-Cloud for mobile based communication has been investigated in detail.  
\par Radio Access Network (RAN) facilitates communication of individual devices with other entities of a network through radio connections. Cloud assisted RAN named CRAN has already drawn significant research attention. In order to complement CRAN, the potentiality of Fog computing based radio access network has also been explored in (Peng et al. \cite{fog_computing_based_radio}).     

\subsection{Long-Reach Passive Optical Network / Power Line Communication}  

Long-Reach Passive Optical Network (LRPON) has been introduced for supporting latency-sensitive and bandwidth-intensive home, industry, and wireless oriented backhaul services. Besides, covering a large area, LRPONs simplify network consolidation process. In (Zhang et al. \cite{infrastructure_deployment_and_optimization}), Fog computing has been integrated with LRPONs for optimized network design. 

\par Power-line communication (PLC) is a widely used communication method in Smart Grid. In PLC, using electrical wiring both data and Alternating Current (AC) are simultaneously transmitted. Fog computing enabled PLC in electric power distribution has been discussed elaborately in (Nazmudeen et al. \cite{improved_throughput_for_power}).

\subsection{Content Distribution Network}

Content Distribution Network (CDN) is composed of distributed proxy servers that provide content to end-users ensuring high performance and availability. In several Fog based research works (Zhu et al. \cite{improving_web_sites_performance}, Do et al. \cite{Obj-carbon_footprint}), Fog nodes are considered as content servers to support content distribution through Fog computing. Since Fog nodes are placed in distributed manner across the edge of the network, Fog based content services can be accessed by the end users within a very minimal delay. As a consequence, high performance in content distribution will be easier to ensure. 

\subsection{Vehicular Network} 

Vehicular network enables autonomous creation of a wireless communication among vehicles for data exchange and resource augmentation. In this networking system vehicles are provided with computational and networking facilities. In several research works (Truong et al. \cite{software_defined_networking_based}, Hou et al. \cite{vehicular_fog_computing_a}, Ye et al. \cite{scalable_fog_computing_with}) vehicles residing at the edge network are considered as Fog nodes to promote Fog computing based vehicular network. 

\section{Security Concern}           

Security vulnerability of Fog computing is very high as it resides at the underlying network between end device/sensors and Cloud datacentres. However, in existing literature, security concerns in Fog computing has been discussed in terms of users authentication, privacy, secured data exchange, DoS attack, etc.

\subsection{Authentication}

Users authentication in Fog based services play an important role in resisting intrusion. Since Fog services are used in "pay as you go" basis, unwanted access to the services are not tolerable in any sense. Besides user authentication, in (Dsouza et al. \cite{policy_driven_security_management}), device authentication, data migration authentication and instance authentication has also been observed for secured Fog computing environment.  

\subsection{Privacy}

Fog computing processes data coming from end device/sensors. In some cases, these data are found very closely associated with users situation and interest. Therefore, proper privacy assurance is considered as one of the important security concerns in Fog computing. In (Hou et al. \cite{vehicular_fog_computing_a}) the challenges regarding privacy in Fog based vehicular computing have been pointed for further investigation.

\subsection{Encryption}

Basically, Fog computing complements Cloud computing. Data that has been processed in Fog computing, in some cases has to be forwarded towards Cloud. As these data often contains sensitive information, it is highly required to encrypt them in Fog nodes. Taking this fact into account, in (Aazam et al. \cite{Archi-Smart_Gateway_Based}), a data encryption layer has been included in the proposed Fog node architecture.

\subsection{DoS Attack}  

Since, Fog nodes are resource constraint, it is very difficult for them to handle large number of concurrent requests. In this case, performance of Fog computing can be degraded to a great extent. To create such severe service disruptions in Fog computing, Denial-of-Service (DoS) attacks can play vital roles. By making a lot of irrelevant service requests simultaneously, Fog nodes can be made busy for a longer period of time. As a result, resources for hosting useful services become unavailable. In (Cirani et al. \cite{Archi-IoT_hub}), this kind of DoS attack in Fog computing has been discussed with clarification.  

\begin{sidewaystable}
\tabfont
\centering
\vspace{11cm} 
\caption{\textbf{Review of state-of-art in Fog Computing}}
\centering
\begin{tabular}{|p{3.1 cm}|p{2.3 cm}|p{2.8 cm}|p{5 cm}|p{3.5 cm}|p{2.6 cm}|p{2.4 cm}p{0 pt}|}
\hline
    \centering \textbf{Work} & \centering \textbf{Fog Nodes} & \centering \textbf{Nodal Collaboration} & \centering \textbf{Provisioning Metrics} & \centering \textbf{SLOs} & \textbf{Applicable Network}  & \textbf{Security Concerns} & \\ \hline 

%
\centering Lee et al.\cite{Archi-WSN} & \centering Servers & \centering Master-Slave &	\centering Data (flow) & \centering Network management &  \centering IoT  &  \centering - &\\
\hline 
%
\centering  Aazam et al.\cite{Model-Micro_Datacenter_Based}& \centering Servers & \centering - & \centering Context (user) & \centering Resource management &  \centering IoT &  \centering - &\\
\hline
%
\centering  Jalali et al.\cite{fog_computing_may_help}& \centering Servers & \centering Peer to Peer & \centering Time (computing) & \centering Power management &  \centering CDN &  \centering - &\\
\hline
%
\centering Zhu et al.\cite{improving_web_sites_performance}& \centering Servers & \centering - & \centering Context (user) & \centering Application management &  \centering CDN &  \centering - &\\
\hline
%
\centering Zeng et al.\cite{joint_optimization_of_task}& \centering Servers & \centering Peer to Peer & \centering Time (communication, computation) & \centering Resource management &  \centering IoT &  \centering - &\\
\hline
%
\centering Hong et al.\cite{mobile_fog_a_programming}& \centering Network devices & \centering Peer to Peer & \centering Data (size) & \centering Application management &  \centering IoT &  \centering - &\\
\hline
%
\centering Nazmudeen et al.\cite{improved_throughput_for_power}& \centering Network devices & \centering Master-Slave & \centering Data (size) & \centering Data management &  \centering PLC &  \centering - &\\
\hline
%
\centering Aazam et al.\cite{Archi-Smart_Gateway_Based} & \centering Network devices & \centering - &	\centering Data (size) & \centering Network management &  \centering IoT  &  \centering Data encryption &\\
\hline 

%
\centering Cirani et al.\cite{Archi-IoT_hub} & \centering Network devices & \centering - &	\centering Context (application) & \centering Network management &  \centering IoT  &  \centering DoS attack &\\
\hline 

%
\centering  Dsouza et al.\cite{policy_driven_security_management}& \centering Cloudlets & \centering Peer to Peer & \centering Context (application) & \centering Network management &  \centering IoT &  \centering Authentication &\\
\hline
%
\centering Cardellini et al.\cite{on_qos_aware_scheduling}& \centering Cloudlets & \centering Cluster & \centering Context (application) & \centering Application management &  \centering IoT &  \centering - &\\
\hline
%
\centering Yan et al.\cite{user_access_mode_selection} & \centering Base stations & \centering Cluster &	\centering Context (user) & \centering Application management &  \centering RAN &  \centering - &\\
\hline 
%
\centering Gu et al.\cite{Obj-Cost_Efficient}& \centering Base stations & \centering Peer to Peer & \centering Cost (deployment, communication) & \centering Resource management &  \centering IoT &  \centering - &\\
\hline
%
\centering Truong et al.\cite{software_defined_networking_based} & \centering Base stations & \centering Master-Slave &	\centering Context (application) & \centering Network management &  \centering Vehicular network &  \centering - &\\
\hline 
%
\centering Oueis et al.\cite{small_cell_clustering_for} & \centering Base stations & \centering Cluster &	\centering Time (deadline) & \centering Latency management &  \centering Mobile network &  \centering - &\\
\hline 
%
\centering Hou et al.\cite{vehicular_fog_computing_a} & \centering Vehicles & \centering Cluster &	\centering Context (user) & \centering Resource management &  \centering Vehicular &  \centering Privacy &\\
\hline 

%
\centering Ye et al.\cite{scalable_fog_computing_with} & \centering Vehicles & \centering - &	\centering Time (deadline) & \centering Application management &  \centering Vehicular &  \centering - &\\
\hline 
%
\centering  Oueis et al.\cite{the_fog_balancing_load}& \centering Base stations & \centering Cluster & \centering Data (flow) & \centering Resource management &  \centering Mobile network &  \centering - &\\
\hline
%
\centering Faruque et al.\cite{energy_management_as_a}& \centering Network devices & \centering Cluster &	\centering Energy consumption & \centering Power management &  \centering IoT &  \centering - &\\
\hline 
%
\centering Shi et al.\cite{combining_mobile_and_fog} & \centering Network devices & \centering Peer to Peer &	\centering Context (application) & \centering Application management &  \centering Mobile network &  \centering - &\\
\hline 

%
\centering  Giang et al.\cite{developing_iot_applications_in}& \centering Network devices & \centering Peer to Peer & \centering Data (flow) & \centering Application management &  \centering IoT &  \centering - &\\
\hline 
%
\centering Intharawijitr et al.\cite{analysis_of_fog_model} & \centering Servers & \centering - &	\centering Time (communication, computation) &	\centering Latency management &  \centering Mobile network  &  \centering - &\\
\hline 
%
\centering Peng et al.\cite{fog_computing_based_radio}& \centering Base stations & \centering Peer to Peer & \centering Data (size) & \centering Network management &  \centering RAN &  \centering - &\\
\hline
%
\centering Hassan et al.\cite{help_your_mobile_applications}& \centering Network devices & \centering Cluster & \centering Cost (execution, communication) & \centering Application management &  \centering Mobile network &  \centering Privacy &\\
\hline
%
\centering  Zhang et al.\cite{infrastructure_deployment_and_optimization}& \centering Cloudlets & \centering Peer to Peer & \centering Cost (deployment) & \centering Cost management &  \centering LRPON &  \centering - &\\
\hline
%
\centering Deng et al.\cite{Obj-Delay_Power}& \centering Network devices & \centering Peer to Peer & \centering Data (Size) & \centering Application management &  \centering Mobile network &  \centering - &\\
\hline
%
\centering Do et al.\cite{Obj-carbon_footprint}& \centering Network devices & \centering - & \centering Energy consumption & \centering $CO_{2}$ management &  \centering CDN &  \centering - &\\
\hline
%
\centering  Aazam et al.\cite{Model-IoT_trace_base}& \centering Servers & \centering - & \centering Context (user) & \centering Resource management &  \centering IoT &  \centering - &\\
\hline
%
\centering Datta et al.\cite{fog_computing_architecture_to}& \centering Network devices & \centering Peer to Peer & \centering Context (user) & \centering Data management &  \centering Vehicular network &  \centering - &\\
\hline
%
\centering Aazam et al.\cite{mefore_qoe_based_resource}& \centering Servers & \centering - & \centering Context (user) & \centering Resource management &  \centering IoT &  \centering - &\\
\hline
%
\centering Gazis et al.\cite{components_of_fog_computing} & \centering Network devices & \centering - &	\centering Context (application) & \centering Resource management &  \centering IoT &  \centering - &\\
\hline 

\end{tabular} 
\end{sidewaystable}

\section{Gap Analysis and Future Directions}
Fog computing resides at closer proximity of the end users and extends Cloud based facilities. In serving largely distributed end devices/sensors, Fog computing plays very crucial roles. Therefore, in recent years Fog computing has become one of the major fields of research from both academia and business perspectives. In Table 1, a brief summary of some reviewed papers from existing literature of Fog computing has been highlighted. 
\par Although many important aspects of Fog computing have been identified in the existing literature, there exist some other issues that are required to be addressed for further improvement of this field. In this section, the gaps from existing literatures along with several future research directions have been discussed.

\subsection{Context-aware Resource/Service Provisioning}

Context-awareness can lead to efficient resource and service provisioning in Fog computing. Contextual information in Fog computing can be received in different forms, for example;
\begin{itemize}
\item Environmental context : Location, Time (Peak, Off-peak), etc.
\item Application context : Latency sensitivity, Application architecture, etc.
\item User context: Mobility, Social interactions, Activity, etc.
\item Device context: Available resources, Remaining battery life time, etc.  
\item Network context: Bandwidth, Network traffic, etc. 
\end{itemize}

Although several Fog based research works have considered contextual information in estimating resources, many important aspects of contextual information are sill unexplored. Investigation of different techniques to apply contextual information in resource and service management can be a potential field towards Fog based research.

\subsection{Sustainable and Reliable Fog computing}

Sustainability in Fog computing optimizes its economic and environmental influence to a great extent. However, the overall sustainable architecture of Fog computing is subject to many issues like assurance of QoS, service reusability, energy-efficient resource management etc. On the other hand, reliability in Fog computing can be discussed in terms of consistency of Fog nods, availability of high performance services, secured interactions, fault tolerance etc. In the existing literature a very narrow discussion towards sustainable and reliable Fog computing has been provided. Further research in this area is highly recommended for the desired performance of Fog computing.

\subsection{Interoperable Architecture of Fog Nodes}

Generally, Fog nodes are specialised networking components with computational facilities. More precisely, besides performing traditional networking activities like packet forwarding, routing, switching, etc., Fog nodes perform computational tasks.  In some scenarios where real time interactions are associated, Fog nodes have to perform more as a computational component rather than a networking component. In other cases, networking capabilities of Fog nodes become prominent over computational capabilities. Therefore, an interoperable architecture of Fog nodes that can be self customized according to the requirements is very necessary. In existing literature although many unique Fog nodes architecture have been proposed, the real interoperable architecture of Fog nodes are still required to be investigated. 

\subsection{Distributed Application Deployment} 

Fog nodes are distributed across the edge and not all of them are highly resource occupied. In this case, large scale application deployment on single Fog node is not often feasible. Modular development of large scale applications and  their distributed deployment over resource constrained Fog nodes can be an effective solution. In existing literature of Fog computing several programming platforms for distribute application development and deployment have been proposed. However, the issues regarding distribute application deployment such as latency management, dataflow management, QoS assurance, edge-centric affinity of real-time applications etc. have not been properly addressed.

\subsection{Power Management within Fog} 

Fog nodes have to deal with huge number of service requests coming from end devices/sensors simultaneously. One of the trivial solutions is to deploy Fog nodes in the environment according to the demand. However, this approach will increase the number of computationally active Fog nodes to a great extent, that eventually affects total power consumption of the system. Therefore, while responding large number service requests, proper power management within Fog network is very necessary. However in existing literature, Fog computing have been considered for minimizing power consumption in Cloud datacentres. Optimization of energy usage within the Fog network are yet to be investigated. Moreover, in order to manage power in Fog environment, consolidation of Fog nodes by migrating tasks from one node to another node can be effective in some scenarios. Investigation towards the solutions of optimal task migration can also be a potential field of Fog based research.  

\subsection{Multi-tenant Support in Fog Resources}

Available resources of Fog nodes can be virtualized and allocated to multiple users. In the existing literature, multi-tenant support in Fog resources and scheduling the computation tasks according to their QoS requirements have not been investigated in detail. Future researches can be conducted targeting this limitation of existing literature.    

\subsection{Pricing, Billing in Fog computing}

Fog computing can provide utility services like Cloud computing. In Cloud computing typically users are charged according to the horizontal scale of usage. Unlike Cloud computing, in Fog vertical arrangement of resources  contributes to the expenses of both users and providers to a great extent. Therefore, the pricing and billing policies in Fog generally differ significantly from the Cloud oriented policies. Besides, due to lack of proper pricing and billing policies of Fog based services, most often users face difficulty in identifying suitable providers for conducting SLA. In such circumstance, a proper pricing and billing policy of Fog based services will surely be considered as potential contribution in field of Fog computing.

\subsection{Tools for Fog Simulation} 
Real-world testbed for evaluating performance of Fog based policies is often very expensive to develop and not scalable in many cases. Therefore, for preliminary evaluation of  proposed Fog computing environments many researchers look for efficient toolkit for Fog simulation. However, till now very less number of Fog simulator are available (e.g. \textit{iFogSim}\cite{ifogsim-Gupta}). Development of new efficient simulator for Fog computing can be taken into account as future research.  

\subsection{Programming Languages and Standards for Fog}

Basically Fog computing has been designed for extending Cloud based services such as IaaS, PaaS, SaaS, etc. to the proximity of IoT devices/sensors. As the structure of Fog differs from Cloud, modification or improvement of existing standards and associate programming languages to enable Cloud-based services in Fog are highly required. Moreover, for seamless and flexible management of large number connections in Fog, development of efficient networking protocols and user interfaces are also necessary.

\section{Summary and Conclusions} 

In this chapter, we surveyed recent developments in Fog computing. Challenges in Fog computing is discussed here in terms of structural, service and security related issues. Based on the identified key challenges and properties, a taxonomy of Fog computing has also been presented. Our taxonomy classifies and analyses the existing works based on their approaches towards addressing the challenges. Moreover, based on the analysis, we proposed some promising research directions that can be pursued in the future.

\bibliographystyle{splncs}
\bibliography{Taxonomy_BookChapter}

\end{document}